%% file: sdpdsl.tex
\documentclass[amsmath,amssymb,pre,twocolumn,showkeys,showpacs,superscriptaddress,preprintnumbers,floatfix,nofootinbib,longbibliography]{revtex4-1}

\usepackage{amsthm}
\usepackage{bbm}
\usepackage{bm}
\usepackage{graphicx}
\usepackage{nicefrac}
\usepackage{url}
\usepackage{verbatim}

\DeclareMathOperator{\E}{E}
\DeclareMathOperator{\Var}{Var}

\input{cmechabbrev}

\input{tikz}
\input{tikz-prerender}
\input{theorems}

\newcommand{\GMP} { \mathrm{GMP} }
\newcommand{\BC}  { \mathrm{BC} }
\newcommand{\FC}  { \mathrm{FC} }
\newcommand{\GM}  { \mathrm{GM} }
\newcommand{\AP}  { \mathrm{AP} }

\begin{document}

\title{Structural Drift:\\
The Population Dynamics of Sequential Learning}

\author{James P. Crutchfield}
\email{chaos@cse.ucdavis.edu}
\affiliation{Complexity Sciences Center}
\affiliation{Physics Department\\
University of California Davis, One Shields Avenue, Davis, CA 95616}
\affiliation{Santa Fe Institute\\
1399 Hyde Park Road, Santa Fe, NM 87501}

\author{Sean Whalen}
\email{shwhalen@gmail.com}
\affiliation{Computer Science Department\\
Columbia University, 1214 Amsterdam Avenue, New York, NY 10027}

\date{\today}

\begin{abstract}
We introduce a theory of sequential causal inference in which learners in a
chain estimate a structural model from their upstream ``teacher'' and then
pass samples from the model to their downstream ``student''. It extends the
population dynamics of genetic drift, recasting Kimura's selectively neutral
theory as a special case of a generalized drift process using structured
populations with memory. We examine the diffusion and fixation properties of
several drift processes and propose applications to learning, inference, and
evolution. We also demonstrate how the organization of drift process space
controls fidelity, facilitates innovations, and leads to information loss in
sequential learning with and without memory.
\end{abstract}

\preprint{Santa Fe Institute Working Paper 10-05-011}
\preprint{arxiv.org:1005.2714 [q-bio.PE]}


\maketitle

\section{``Send Three- and Four-Pence, We're Going to a Dance''}

This phrase was heard, it is claimed, over the radio during WWI instead of the
transmitted tactical phrase ``Send reinforcements we're going to advance''
\cite{Smith1988}. As illustrative as it is apocryphal, this garbled yet
comprehensible transmission sets the tone for our investigations here. Namely,
what happens to knowledge when it is communicated sequentially along a chain,
from one individual to the next? What fidelity can one expect? How is
information lost? How do innovations occur?

To answer these questions we introduce a theory of sequential causal inference
in which learners in a communication chain estimate a structural model from
their upstream ``teacher'' and then, using that model, pass along samples to
their downstream ``student''. This reminds one of the familiar children's game
\emph{Telephone}. By way of quickly motivating our sequential learning problem,
let's briefly recall how the game works.

To begin, one player invents a phrase and whispers it to another player. This
player, believing they have understood the phrase, then repeats it to a third
and so on until the last player is reached. The last player announces the
phrase, winning the game if it matches the original. Typically it does not, and
that's the fun. Amusement and interest in the game derive directly from how the
initial phrase evolves in odd and surprising ways. The further down the chain,
the higher the chance that errors will make recovery impossible and the less
likely the original phrase will survive.

The game is often used in education to teach the lesson that human
communication is fraught with error. The final phrase, though, is not merely
accreted error but the product of a series of attempts to parse, make sense,
and intelligibly communicate the phrase. The phrase's evolution is a trade off
between comprehensibility and accumulated distortion, as well as the source of
the game's entertainment. We employ a much more tractable setting to make
analytical progress on sequential learning, based on \emph{computational
mechanics} \cite{Crutchfield1989,Crutchfield1992,Shalizi2001a}, intentionally
selecting a simpler language system and learning paradigm than likely operates
with children.

Specifically, we develop our theory of sequential learning as an extension of
the evolutionary population dynamics of genetic drift, recasting Kimura's
selectively neutral theory \cite{Kimura1969} as a special case of a generalized
drift process of structured populations with memory. This is a substantial
departure from the unordered populations used in evolutionary biology. Notably,
this requires a new and more general information-theoretic notion of fixation.
We examine the diffusion and fixation properties of several drift processes,
demonstrating that the space of drift processes is highly organized. This
organization controls fidelity, facilitates innovations, and leads to
information loss in sequential learning and evolutionary processes with and
without memory. We close by describing applications to learning, inference, and
evolution, commenting on related efforts.

To get started, we briefly review genetic drift and fixation. This will seem
like a distraction, but it is a necessary one since available mathematical
results are key. Then we introduce in detail our structured variants of these
concepts---defining the \emph{generalized drift process} and formulating a
generalized definition of fixation appropriate to it. With the background laid
out, we begin to examine the complexity of structural drift behavior. We
demonstrate that it is a diffusion process within a space that decomposes into
a connected network of structured subspaces. Building on this decomposition, we
explain how and when processes jump between these subspaces---innovating new
structural information or forgetting it---thereby controlling the long-time
fidelity of the communication chain. We then close by outlining future research
and listing several potential applications for structural drift, drawing out
consequences for evolutionary processes that learn.

Those familiar with neutral evolution theory are urged to skip to Sec.
\ref{section:sequential_learning}, after skimming the next sections to pick up
our notation and extensions.

\section{From Genetic to Structural Drift}

Genetic drift refers to the change over time in genotype frequencies in a
population due to random sampling. It is a central and well studied phenomenon
in population dynamics, genetics, and evolution. A population of genotypes
evolves randomly due to drift, but typically changes are neither manifested as
new phenotypes nor detected by selection---they are \emph{selectively neutral}.
Drift plays an important role in the spontaneous emergence of mutational
robustness \cite{VanNimwegen1999,Bloom2006}, modern techniques for calibrating
molecular evolutionary clocks \cite{Raval2007}, and nonadaptive (neutral)
evolution \cite{Crutchfield2003a,Koelle2006}, to mention only a few examples.

Selectively neutral drift is typically modeled as a stochastic process: A
random walk that tracks finite populations of individuals in terms of their
possessing (or not) a variant of a gene. In the simplest models, the random
walk occurs in a space that is a function of genotypes in the population. For
example, a drift process can be considered to be a random walk of the
\emph{fraction} of individuals with a given variant. In the simplest cases
there, the model reduces to the dynamics of repeated binomial sampling of a
biased coin, in which the empirical estimate of bias becomes the bias in the
next round of sampling. In the sense we will use the term, the sampling process
is \emph{memoryless}. The biased coin, as the population being sampled, has no
memory: The past is independent of the future. The current state of the drift
process is simply the bias, a number between zero and one that summarizes the
state of the population.

The theory of genetic drift predicts a number of measurable properties. For
example, one can calculate the expected time until all or no members of a
population possess a particular gene variant. These final states are referred
to as \emph{fixation} and \emph{deletion}, respectively. Variation due to
sampling vanishes once these states are reached and, for all practical
purposes, drift stops. From then on, the population is homogeneous; further
sampling can introduce no genotypic variation. These states are fixed
points---in fact, absorbing states---of the drift stochastic process.

The analytical predictions for the time to fixation and time to deletion were
developed by Kimura and Ohta \cite{Kimura1969,Kimura1983} in the 1960s and are
based on the memoryless models and simplifying assumptions introduced by Wright
\cite{Wright1931} and Fisher \cite{Fisher1930} in the early 1930s. The theory
has advanced substantially since then to handle more realistic models and to
predict additional effects due to selection and mutation. These range from
multi-allele drift models and $F$-statistics \cite{Holsinger2009} to
pseudohitchhiking models of ``genetic draft'' \cite{Gillespie2000}.

The following explores what happens when we relax the memoryless restriction.
The original random walk model of genetic drift forces the statistical
structure at each sampling step to be an independent, identically distributed
(IID) stochastic process. This precludes any memory in the sampling. Here, we
extend the IID theory to use time-varying probabilistic state machines to
describe memoryful population sampling.

In the larger setting of sequential learning, we will show that memoryful
sequential sampling exhibits structurally complex, drift-like behavior. We call
the resulting phenomenon \emph{structural drift}. Our extension presents a
number of new questions regarding the organization of the space of drift
processes and how they balance structure and randomness. To examine these
questions, we require a more precise description of the original drift theory.

\section{Genetic Drift}

We begin with the definition of an \emph{allele}, which is one of several
alternate forms of a gene. The textbook example is given by Mendel's early
experiments on heredity \cite{Mendel1925}, in which he observed that the
flowers of a pea plant were colored either white or violet, this being
determined by the combination of alleles inherited from its parents. A new,
\emph{mutant} allele is introduced into a population by the mutation of a
\emph{wild-type} allele. A mutant allele can be passed on to an individual's
offspring who, in turn, may pass it on to their offspring. Each inheritance
occurs with some probability.

\emph{Genetic drift}, then, is the change of allele frequencies in a
population over time. It is the process by which the number of individuals
with an allele varies generation after generation. The Fisher-Wright theory
\cite{Wright1931,Fisher1930} models drift as a stochastic evolutionary process
with neither selection nor mutation. It assumes random mating between
individuals and that the population is held at a finite, constant size.
Moreover, successive populations do not overlap in time.

Under these assumptions the Fisher-Wright theory reduces drift to a binomial
or multinomial sampling process---a more complicated version of familiar
random walks such as Gambler's Ruin or Prisoner's Escape \cite{Feller1968}.
Offspring receive either the wild-type allele $A_1$ or the mutant allele $A_2$
of a particular gene $\MeasAlphabet$ from a random parent in the previous
generation with replacement. A population of $N$ diploid\footnotemark\
individuals will have $2N$ total copies of these alleles. Given $i$ initial
copies of $A_2$ in the population, an individual has either $A_2$ with
probability $\nicefrac{i}{2N}$ or $A_1$ with probability $1-\nicefrac{i}{2N}$.
The probability that $j$ copies of $A_2$ exist in the offspring's generation
given $i$ copies in the parent's generation is:
\begin{align}
p_{ij} = \binom{2N}{j} \left(\frac{i}{2N}\right)^j
\left(1-\frac{i}{2N}\right)^{2N-j}.
\end{align}
This specifies the transition dynamic of the drift stochastic process over the
discrete state space $\{0, \nicefrac{1}{N}, \ldots, \nicefrac{N-1}{N}, 1\}$.

\setcounter{footnote}{0}
\footnotetext{Though we first use diploid populations (two alleles per
individual and thus a sample length of $2N$) for direct comparison to previous
work, we later transition to haploid (single allele per
individual) populations for notational simplicity.}

This model of genetic drift is a discrete-time random walk, driven by samples
of a biased coin, over the space of biases. The population is a set of coin
flips, where the probability of \textsc{Heads} or \textsc{Tails} is determined
by the coin's current bias. After each generation of flips, the coin's bias is
updated to reflect the number of \textsc{Heads} or \textsc{Tails} realized in
the new generation. The walk's absorbing states---all \textsc{Heads} or all
\textsc{Tails}---capture the notion of fixation and deletion.

\section{Genetic Fixation}

\emph{Fixation} occurs with respect to an allele when all individuals in the
population carry that specific allele and none of its variants. Restated, a
mutant allele $A_2$ reaches fixation when all ${2N}$ alleles in the population
are copies of $A_2$ and, consequently, $A_1$ has been {\em deleted} from the
population. This halts the random fluctuations in the frequency of $A_2$,
assuming $A_1$ is not reintroduced.

Let $X$ be a binomially distributed random variable with bias probability $p$
that represents the fraction of copies of $A_2$ in the population. The
expected number of copies of $A_2$ is $\E[X] = 2Np$. That is, the expected
number of copies of $A_2$ remains constant over time and depends only on its
initial probability $p$ and the total number ($2N$) of alleles in the
population. However, $A_2$ eventually reaches fixation or deletion due to the
change in allele frequency introduced by random sampling and the presence of
absorbing states. Prior to fixation, the mean and variance of the change in
allele frequency $\Delta p$ are:
\begin{align}
\E[\Delta p] &= 0 \mathrm{\ and}\\
\Var[\Delta p] &= \frac{p\,(1-p)}{2N},
\end{align}
respectively.

On average there is no change in frequency. However, sampling variance causes
the process to drift towards the absorbing states at \mbox{$p = 0$} and
\mbox{$p = 1$}. The drift rate is determined by the current generation's
allele frequency and the total number of alleles. For the neutrally selective
case, the average number of generations until fixation ($t_1$) or deletion
($t_0$) is given by Kimura and Ohta \cite{Kimura1969}:
\begin{align}
t_1(p) &= -\frac{1}{p}\left[ 4N_e(1-p)\log (1-p)\right] \mathrm{\ and}
  \label{eq:time_to_fixation} \\
t_0(p) &= -4N_e \left( \frac{p}{1-p} \right) \log p ~,
  \label{eq:time_to_deletion}
\end{align}
where $N_e$ denotes effective population size. For simplicity we take $N_e =
N$, meaning all individuals in the population are candidates for reproduction.
As \mbox{$p \to 0$}, the boundary condition is given by:
\begin{align}
t_1(0) = 4N_e ~.
\end{align}
That is, excluding cases of deletion, an initially rare mutant allele spreads
to the entire population in $4N_e$ generations.

\begin{figure*}
\begin{minipage}[t]{\columnwidth}
\centering
\includegraphics{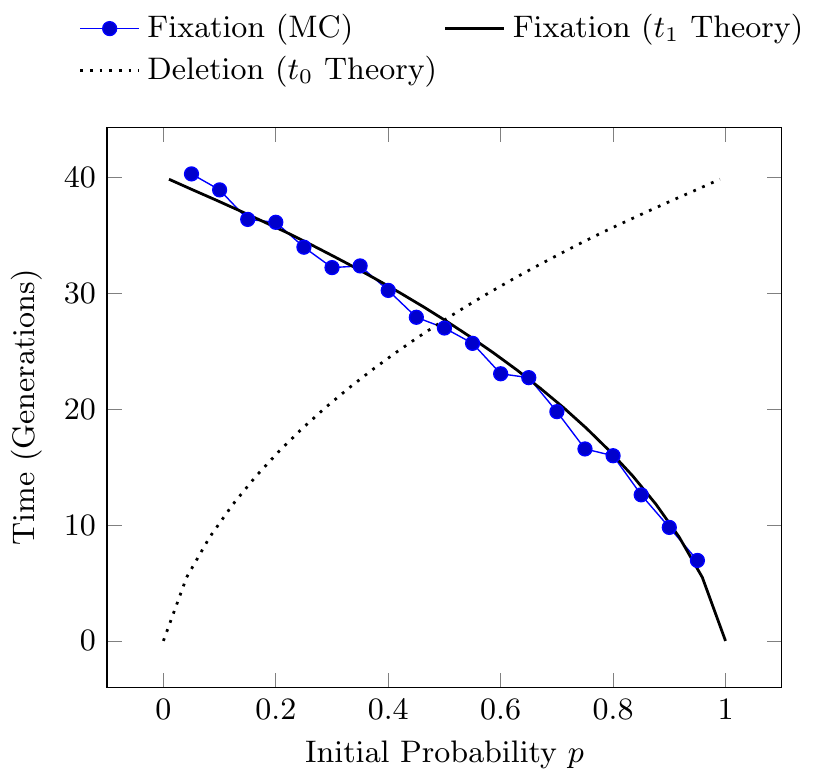}
\caption{
Time to fixation for a population of \mbox{$N = 10$} individuals (sample size
$2N = 20$) plotted as a function of initial allele probability $p$ under the
Monte Carlo (MC) sampling regime and as given by theoretical prediction (solid
line) of Eq. (\ref{eq:time_to_fixation}). Time to deletion is also
shown (dashed line), Eq. (\ref{eq:time_to_deletion}).
}
\label{figure:fixation_kimura}
\end{minipage}
\hspace{\fill}
\begin{minipage}[t]{\columnwidth}
\centering
\includegraphics{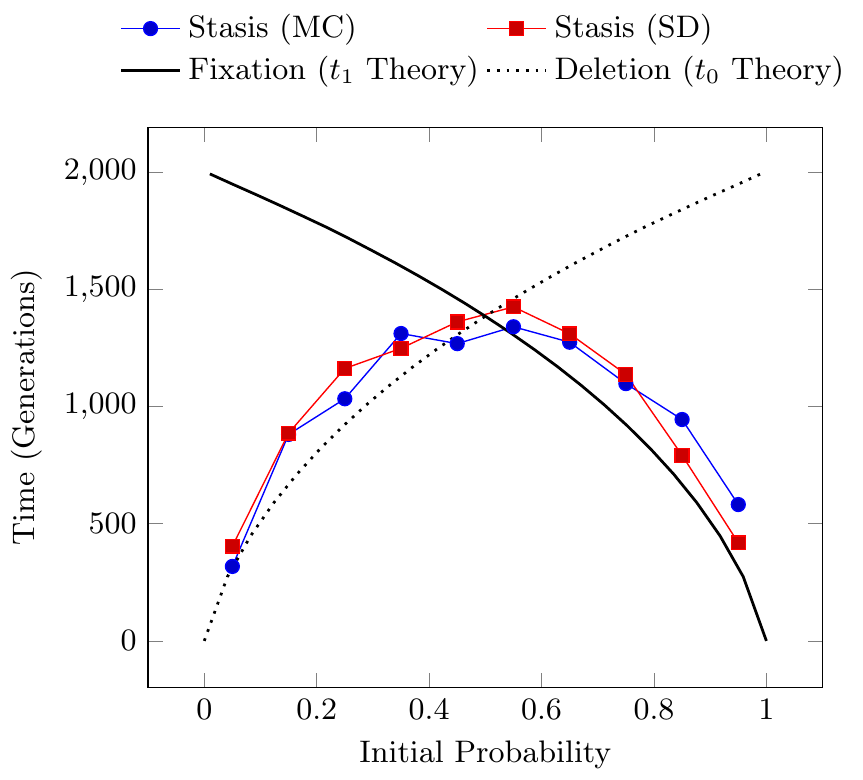}
\caption{
Time to stasis as a function of initial $\Prob[\textsc{Heads}]$ for structural
drift (SD) of the Biased Coin Process versus Monte Carlo (MC) simulation of
Kimura's model. Kimura's predicted times to fixation and deletion are shown
for reference. Each estimated time is averaged over $100$ realizations with
sample size \mbox{$N = 1000$}.
}
\label{figure:tts_bc}
\end{minipage}
\end{figure*}

One important consequence of the theory is that when fixation (\mbox{$p = 1$})
or deletion (\mbox{$p = 0$}) are reached, variation in the population vanishes:
\mbox{$\Var[\Delta p] = 0$}. With no variation there is a homogeneous
population, and sampling from this population produces the same homogeneous
population. In other words, this establishes fixation and deletion as absorbing
states of the stochastic sampling process. Once there, drift stops.

Figure \ref{figure:fixation_kimura} illustrates this, showing both the
simulated and theoretically predicted number of generations until fixation
occurs for \mbox{$N = 10$}, as well as the predicted time to deletion for
reference. Each simulation was performed for a different initial value of $p$
and averaged over 400 realizations. Using the same methodology as Kimura and
Ohta \cite{Kimura1969}, we include only those realizations whose mutant allele
reaches fixation.

Populations are produced by repeated binomial sampling of $2N$ uniform random
numbers between $0$ and $1$. An initial probability \mbox{$1 - p$} is assigned
to allele $A_1$ and probability $p$ to allele $A_2$. The count $i$ of $A_2$ in
the initial population is incremented for each random number less than $p$.
This represents an individual acquiring the allele $A_2$ instead of $A_1$. The
maximum likelihood estimate of allele frequency in the initial sample is simply
the number of $A_2$ alleles over the sample length: \mbox{$p =
\nicefrac{i}{2N}$}. This estimate of $p$ is then used to generate a new
population of offspring, after which we re-estimate the value of $p$. These
steps are repeated each generation until fixation at \mbox{$p = 1$} or deletion
at \mbox{$p = 0$} occurs. This is the \emph{Monte Carlo} (MC) sampling method.

Kimura's theory and simulations predict the time to fixation or deletion of a
mutant allele in a finite population by the process of genetic drift. The
Fisher-Wright model and Kimura's theory assume a memoryless population in which
each offspring inherits allele $A_1$ or $A_2$ via an IID binomial sampling
process. We now generalize this to memoryful stochastic processes, giving a new
definition of fixation and exploring examples of structural drift behavior.

\section{Sequential Learning}
\label{section:sequential_learning}

How can genetic drift be a memoryful stochastic process? Consider a population
of $N$ haploid organisms. Each generation consists of $N$ alleles and so is
represented by a string of $N$ symbols, e.g. $A_1 A_2 \ldots A_1 A_1$, where
each symbol corresponds to an individual with a particular allele. In the
original drift models, a generation of offspring is produced by a memoryless
binomial sampling process, selecting an offspring's allele from a parent with
replacement. In contrast, the structural drift model produces a generation of
individuals in which the sample order is tracked. The population is now a
string of alleles, giving the potential for memory and structure in
sampling---spatial, temporal, or other interdependencies between individuals
within a sample.

At first, this appears as a major difference from the usual setting employed in
evolutionary biology, where populations are treated as unordered collections of
individuals and sampling is modeled as an independent, identically distributed
stochastic process. That said, the structure we have in mind has several
biological interpretations, such as inbreeding and subdivision
\cite{Gillespie2004} or the life histories of heterogeneous populations
\cite{Leibler2010}. We later return to these alternative interpretations when
considering applications.

The model class we select to describe memoryful sampling is the \eM: the
unique, minimal, and optimal representation of a stochastic process
\cite{Shalizi2001a}. As we will show, these properties give an important
advantage when analyzing structural drift, since they allow one to monitor the
amount of structure innovated or lost during drift. We next give a brief
overview of \eMs\ and refer the reader to the previous reference for details.

\EM\ representations of the finite-memory discrete-valued stochastic processes
we consider here form a class of (deterministic) probabilistic finite-state
machine or unifilar hidden Markov model. An \eM\ consists of a set of
\emph{causal states} \mbox{$\CausalStateSet = \{0, 1, \ldots, k-1\}$} and a set
of per-symbol transition matrices:
\begin{align}
\{T_{ij}^{(a)}: a \in \MeasAlphabet \} ~,
\end{align}
where \mbox{$\MeasAlphabet = \{A_1, \ldots, A_m \}$} is the set of alleles and
where the transition probability $T_{ij}^{(a)}$ gives the probability of
transitioning from causal state $\CausalState_i$ to causal state
$\CausalState_j$ and emitting allele $a$. The causal state probability
$\Prob(\sigma)$, $\sigma \in \CausalStateSet$, is determined as the left
eigenvector of the state-to-state transition matrix $T = \sum_{a \in
\MeasAlphabet} T^{(a)}$.

Maintaining our connection to (haploid) population dynamics, we think of an
\eM\ as a generator of populations or length-$N$ strings: \mbox{$\alpha^{N} =
a_1 a_2 \ldots a_i \ldots a_{N}, a_i \in \MeasAlphabet$}. As a model of a
sampling process, an \eM\ gives the most compact representation of the
distribution of strings produced by sampling.

\begin{figure}
\begin{minipage}[t]{\columnwidth}
\centering
\includegraphics{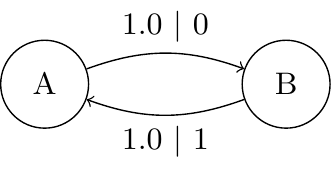}
\caption{
\EM\ for the Alternating Process consisting of two causal states
\mbox{$\CausalStateSet = \{A, B\}$} and two transitions. Each transition is
labeled $p\,|\,a$ to indicate the probability \mbox{$p = T_{ij}^{(a)}$} of
taking that transition and emitting allele \mbox{$a \in \MeasAlphabet$}. State
$A$ emits allele 0 with probability one and transitions to state $B$, while
$B$ emits allele 1 with probability one and transitions to $A$.
}
\label{figure:em_period2}
\end{minipage}
\hspace{\fill}
\begin{minipage}[b]{\columnwidth}
\centering
\includegraphics{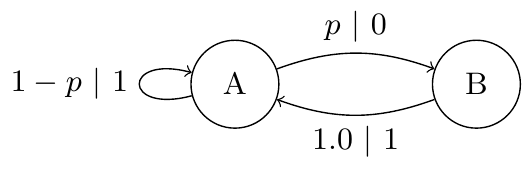}
\caption{
The \eM\ for the Golden Mean Process that generates a population with no
consecutive $0$s. In state $A$ the probabilities of generating a 0 or 1 are $p$
and $1-p$, respectively.
}
\label{figure:em_gm}
\end{minipage}
\end{figure}

Consider a simple binary process that alternately generates $0$s and $1$s
called the \emph{Alternating Process} shown in Fig. \ref{figure:em_period2}.
Its \eM\ generates either the string $0101 \dots$ or $1010 \dots$ depending on
the start state. The per-symbol transition matrices are:
\begin{align}
    T^{(0)} &= \begin{pmatrix} 0.0 & 1.0 \\ 0.0 & 0.0 \end{pmatrix}
\mathrm{\ and}\\
    T^{(1)} &= \begin{pmatrix} 0.0 & 0.0 \\ 1.0 & 0.0 \end{pmatrix}.
\end{align}
Enforcing the alternating period-2 pattern requires two states, $A$ and $B$,
as well as two positive probability transitions \mbox{$T_{AB}^{(0)} = 1.0$}
and \mbox{$T_{BA}^{(1)} = 1.0$}.

\begin{figure}[b]
\includegraphics{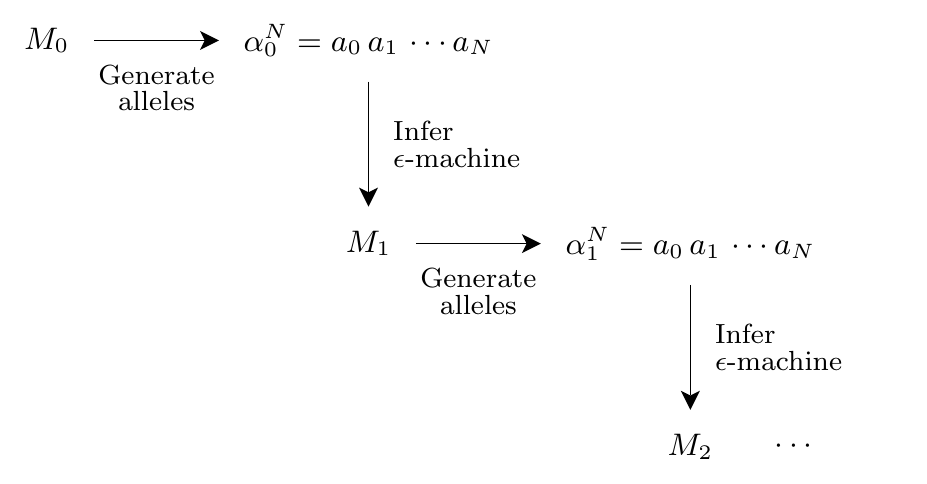}
\caption{
Sequential inference of a chain of \eMs. An initial population generator
$\eMachine_0$ produces a length-$N$ string $\alpha^{N}_0 = a_1 \ldots a_{N}$
from which a new model $\eMachine_1$ is inferred. These steps are repeated
using $\eMachine_1$ as the population generator and so on, until a terminating
condition is met.
}
\label{figure:sequential_inference}
\end{figure}

We are now ready to describe \emph{sequential learning}, depicted in Fig.
\ref{figure:sequential_inference}. We begin by selecting an initial population
generator $\eMachine_0$---an \eM. Following a path through $\eMachine_0$,
guided by its transition probabilities, produces a length-$N$ string
$\alpha^{N}_0 = a_1 \ldots a_{N}$ that represents the first population of $N$
individuals possessing alleles $a_i \in \MeasAlphabet$. We then infer an \eM\
$\eMachine_1$ from the population $\alpha^{N}_0$. $\eMachine_1$ is then used
to produce a new population $\alpha^{N}_1$, from which a new \eM\
$\eMachine_2$ is estimated. This new population has the same allele
distribution as the previous, plus some amount of variance. The cycle of
inference and re-inference is repeated while allele frequencies drift each
generation until fixation or deletion is reached. At that point, the
populations (and so \eMs) cannot vary further. The net result is a
stochastically varying time series of \eMs\ ($\eMachine_0, \eMachine_1,
\eMachine_2, \ldots$) that terminates when the populations $\alpha_t^{N}$
stop changing.

Thus, at each step a new representation or model is estimated from the
previous step's sample. The inference step highlights that this is learning: a
model of the generator is estimated from the given finite data. The repetition
of this step creates a sequential communication chain. Sequential learning is
thus closely related to genetic drift except that sample order is tracked, and
this order is used in estimating the next model.

The procedure is analogous to flipping a biased coin a number of times,
estimating the bias from the results, and re-flipping the newly biased coin.
Eventually, the coin will be completely biased towards \textsc{Heads} or
\textsc{Tails}. In our drift model the coin is replaced by an \eM, which
removes the IID model constraint and allows for the sampling process to take
on structure and memory. Not only do the transition probabilities
$T^{(a)}_{ij}$ change, but \emph{the structure of the model itself}---the
number of states and the presence or absence of transitions---drifts over time
to capture the statistics of the sample using as little information as
possible. This is an essential and distinctive aspect of structural drift.

Before we can explore this dynamic, we first need to examine how an \eM\
reaches fixation or deletion.

\section{Structural Stasis}

Recall the Alternating Process from Fig. \ref{figure:em_period2}, producing
the strings $0101 \dots$ and $1010 \dots$ depending on the start state.
Regardless of the initial state, the original \eM\ is re-inferred from any
sufficiently long string it produces. In the context of sequential learning,
this means the population at each generation is the same.

However, if we consider allele $A_1$ to be represented by symbol $0$ and $A_2$
by symbol $1$, neither allele reaches fixation or deletion according to current
definitions. Nonetheless, the Alternating Process prevents any variance between
generations and so, despite the population not being all $0$s or all $1$s, the
population does reach an equilibrium: half $0$s and half $1$s. For these
reasons, one cannot use the original population-dynamics definitions of
fixation and deletion.

This leads us to introduce \emph{structural stasis} to combine the notions of
fixation, deletion, and the inability to vary caused by periodicity. Said more
directly, structural stasis corresponds to a process becoming nonstochastic,
since it ceases to introduce variance between generations and so prevents
further drift. However, we need a method to detect the occurrence of
structural stasis in a drift process.

A state machine representing a periodic sampling process enforces the
constraint of periodicity via its internal memory. One measure of this memory
is the \emph{population diversity} $H(N)$~\cite{Pielou1967}:
\begin{align}
H(N) &= H[\MeasAlphabet_1 \dots \MeasAlphabet_{N}] \\
  &= -\sum_{a^{N} \in \mathcal{A}^{N}} \Prob(a^{N}) \log_2 \Prob(a^{N}) ~,
\end{align}
where the units are [bits].\footnotemark\ The population diversity of the
Alternating Process is \mbox{$H(N) = 1$} bit at any size \mbox{$N \gg 1$}.
This single bit of information corresponds to the machine's current phase or
state. Generally, though, the value diverges---$H(N) \propto N$---for
arbitrary sampling processes, and so population diversity is not suitable as a
general test for stasis.

\footnotetext{For background on information theory as used here, the reader
is referred to Ref. \cite{Crutchfield2003}.}

\begin{figure*}
\begin{minipage}[t]{\columnwidth}
\centering
\includegraphics{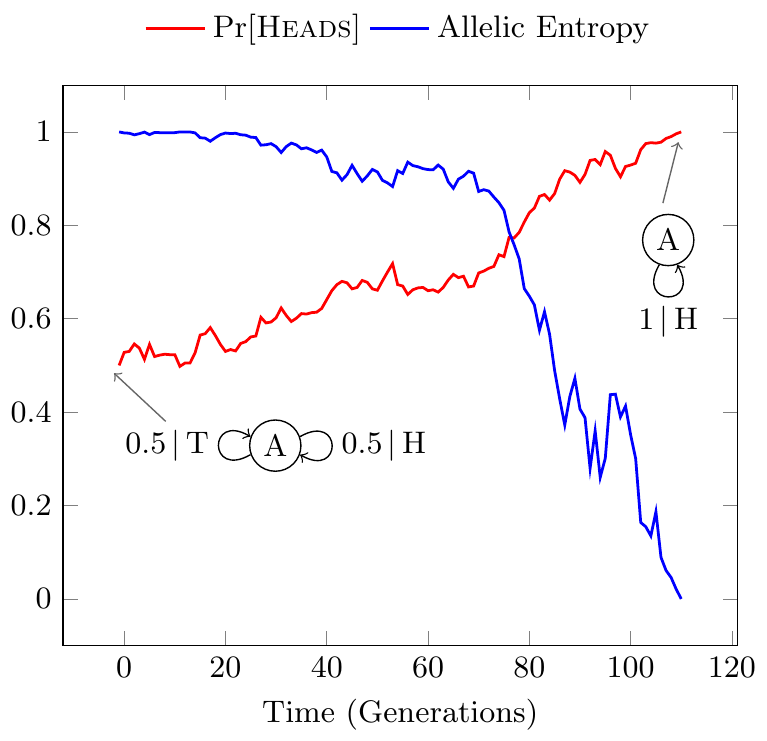}
\caption{
Drift of allelic entropy $\hmu$ and $\Prob[\textsc{Heads}]$ for a single
realization of the Biased Coin Process with sample size~\mbox{$N = 100$}.
The drift of $\Prob[\textsc{Heads}]$ is annotated with its initial machine
$\eMachine_0$ (left inset) and the machine at stasis $\eMachine_{115}$ (right
inset).
}
\label{figure:drift_bc}
\end{minipage}
\hspace{\fill}
\begin{minipage}[t]{\columnwidth}
\centering
\includegraphics{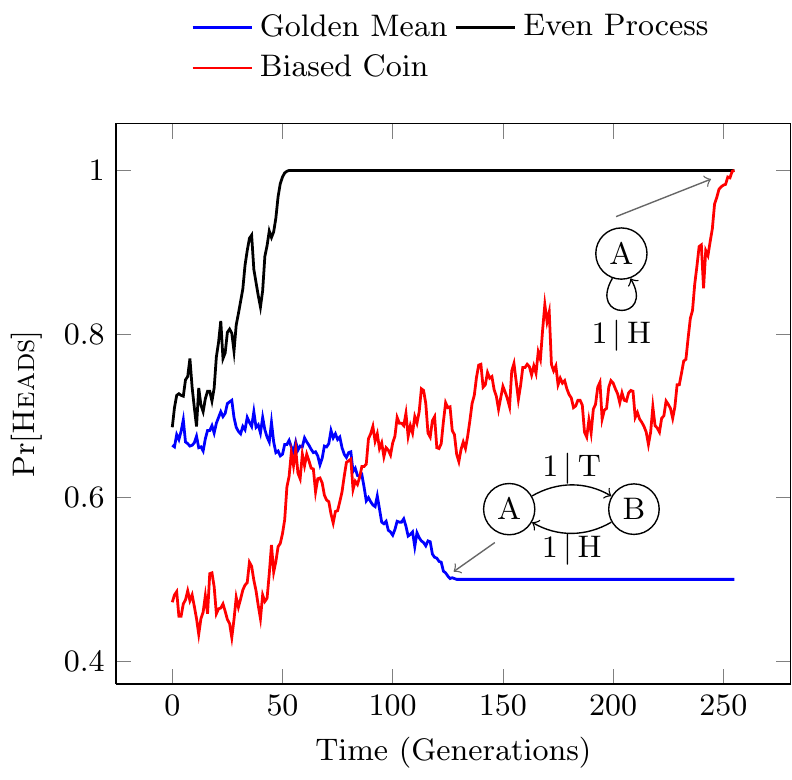}
\caption{
Drift of $\Prob[\textsc{Heads}]$ for a single realization of the Biased Coin,
Golden Mean, and Even Processes, plotted as a function of generation. The Even
and Biased Coin Processes become the Fixed Coin Process at stasis, while the
Golden Mean Process becomes the Alternating Process. Note that the definition
of structural stasis recognizes the lack of variance in the Alternating
Process subspace even though the allele probability is neither 0 nor 1.
}
\label{figure:drift_multiple}
\end{minipage}
\end{figure*}

Instead, the condition for stasis can be given as the vanishing of the
\emph{growth rate} of population diversity:
\begin{align}
\hmu = \lim_{N \to \infty} \bigl[H(N) - H(N-1)\bigr] .
\end{align}
Equivalently, we can test the per-allele entropy of the sampling process. We
call this \emph{allelic entropy}:
\begin{align}
\hmu = \lim_{N \to \infty} \frac{H(N)}{N} ~,
\end{align}
where the units are [bits per allele]. Allelic entropy gives the average
information per allele in bits, and structural stasis occurs when \mbox{$\hmu
= 0$}. While closer to a general test for stasis, this quantity is difficult
to estimate from population samples since it relies on an asymptotic estimate
of the population diversity. However, the allelic entropy can be calculated in
closed-form from the \eM\ representation of the sampling process:
\begin{align}
\hmu & = -\sum_{\causalstate \in \CausalStateSet}
  \Prob(\causalstate)
  \sum_{ \substack{a \in \MeasAlphabet \\
                  \causalstate^\prime \in \CausalStateSet}
       }
  T^{(a)}_{\causalstate\causalstate^\prime}
  \log_2
  T^{(a)}_{\causalstate\causalstate^\prime}
  ~,
\end{align}
When $\hmu = 0$, the sampling process has become periodic and lost all
randomness generated via its branching transitions. This new criterion
subsumes the notions of fixation and deletion as well as periodicity. An \eM\
has zero allelic entropy if any of these conditions occur. More formally, we
have the following statement.

\begin{Def}
\emph{Structural stasis} occurs when the sampling process's allelic entropy
vanishes: $\hmu = 0$.
\end{Def}

\begin{Prop}
Structural stasis is a fixed point of finite-memory structural drift.
\end{Prop}

\begin{ProProp}
Finite-memory means that the \eM\ representing the population sampling process
has a finite number of states. Given this, if \mbox{$\hmu = 0$}, then the \eM\
has no branching in its recurrent states: \mbox{$T^{(a)}_{ij} = 0\
\mathrm{or}\ 1$}, where $\CausalState_i$ and $\CausalState_j$ are
asymptotically recurrent states. This results in no variation in the inferred
\eM\ when sampling sufficiently large populations. Lack of variation, in turn,
means that $\Delta p = 0$ and so the drift process stops. If allelic entropy
vanishes at time $t$ and no mutations are allowed, then it is zero for all
\mbox{$t^\prime > t$}. Thus, structural stasis is an absorbing state of the
drift stochastic process.
\end{ProProp}

\section{Examples}

While more can be said analytically about structural drift, our present purpose
is to introduce the main concepts. We will show that structural drift leads to
interesting and nontrivial behavior. First, we calibrate the new class of drift
processes against the original genetic drift theory.

\subsection{Memoryless Drift}

The Biased Coin Process is represented by a single-state \eM\ with a self loop
for both \textsc{Heads} and \textsc{Tails} symbols. It is an IID sampling
process that generates populations with a binomial distribution. Unlike the
Alternating Process, the coin's bias $p$ is free to drift during sequential
inference. These properties make the Biased Coin Process an ideal candidate for
exploring memoryless drift.

Fig. \ref{figure:drift_bc} shows structural drift, using two different
measures, for a single realization of the Biased Coin Process with initial
\mbox{$p = \Prob[\textsc{Heads}] = \Prob[\textsc{Tails}] = 0.5$}. Structural
stasis ($\hmu = 0)$ is reached after $115$ generations. The initial Fair Coin
\eM\ occurs at the left of Fig. \ref{figure:drift_bc} and the final, completely
biased \eM\ occurs at the right.

Note that the drift of allelic entropy $\hmu$ and \mbox{$p =
\Prob[\textsc{Tails}]$} are inversely related, with allelic entropy converging
quickly to zero as stasis is approached. This reflects the rapid drop in
population diversity. After stasis occurs, all randomness has been eliminated
from the transitions at state $A$, resulting in a single transition that always
produces \textsc{Tails}. Anticipating later discussion, we note that during
this run only Biased Coin Processes were observed.

The time to stasis of the Biased Coin Process as a function of initial \mbox{$p
= \Prob[\textsc{Heads}]$} was shown in Fig. \ref{figure:tts_bc}. Also shown
there was the previous Monte Carlo Kimura drift simulation modified to
terminate when either fixation or deletion occurs. This experiment illustrates
the definition of structural stasis and allows direct comparison of structural
drift with genetic drift in the memoryless case.

Not surprisingly, we can interpret genetic drift as a special case of the
structural drift process for the Biased Coin. Both simulations follow Kimura's
theoretically predicted curves, combining the lower half of the deletion curve
with the upper half of the fixation curve to reflect the initial probability's
proximity to the absorbing states. A high or low initial bias leads to a
shorter time to stasis as the absorbing states are closer to the initial state.
Similarly, a Fair Coin is the furthest from absorption and thus takes the
longest average time to reach stasis.

\subsection{Structural Drift}

The Biased Coin Process represents an IID sampling process with no memory of
previous flips, reaching stasis when \mbox{$\Prob[\textsc{Heads}] = 1.0$} or
0.0 and, correspondingly, when \mbox{$\hmu(\eMachine_t) = 0.0$}. We now
introduce memory by starting drift with $\eMachine_0$ as the \emph{Golden Mean
Process}, which produces binary populations with no consecutive $0$s. Its \eM\
was shown in Fig. \ref{figure:em_gm}. Note that one can initialize drift using
any stochastic process; for example, see the \eM\ library of Ref.
\cite{Johnson2010}.

Like the Alternating Process, the Golden Mean Process has two causal states.
However, the transitions from state $A$ have nonzero entropy, allowing their
probabilities to drift as new \eMs\ are inferred from generation to
generation. If the \mbox{$A \to B$} transition probability~$p$ (Fig.
\ref{figure:em_gm}) becomes zero the transition is removed, and the Golden
Mean Process reaches stasis by transforming into the Fixed Coin Process (top
right, Fig. \ref{figure:drift_bc}). Instead, if the same transition drifts
towards probability \mbox{$p = 1$}, the \mbox{$A \to A$} transition is
removed. In this case, the Golden Mean Process reaches stasis by transforming
into the Alternating Process (Fig. \ref{figure:em_period2}).

\begin{figure*}
\begin{minipage}[t]{\columnwidth}
\centering
\includegraphics{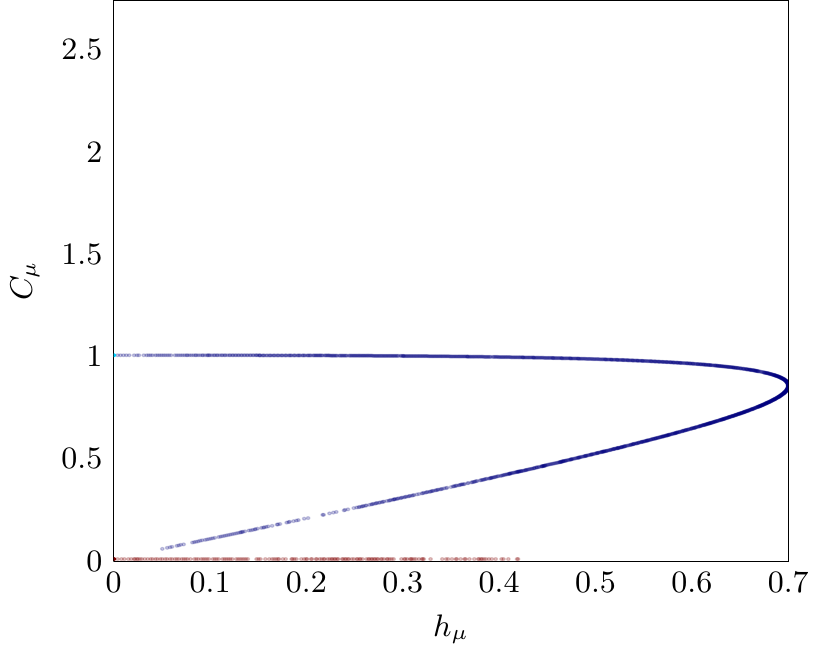}
\end{minipage}
\hspace{\fill}
\begin{minipage}[t]{\columnwidth}
\centering
\includegraphics{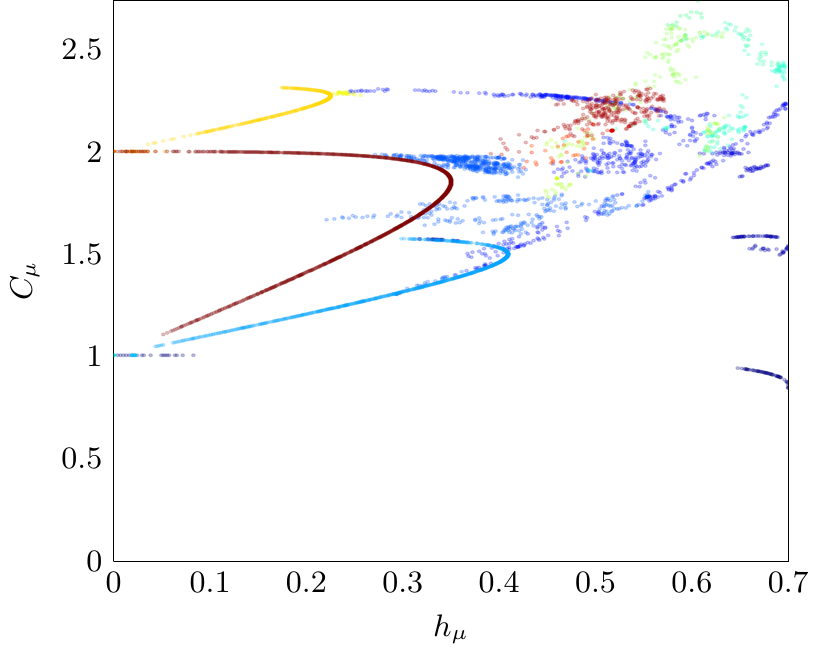}
\end{minipage}
\caption{
Complexity-entropy diagram for $30$ realizations of the Golden Mean
Process with \mbox{$N = 1000$}, both without (left) and with (right)
structural innovation. Alternating Process and Fixed Coin pathways are clearly
visible in the left panel where the Golden Mean subspace exists on the upper
curve and the Biased Coin subspace exists on the line \mbox{$\Cmu = 0$}. \EMs\
within the same isostructural subspace have identical colors.
}
\label{figure:ce_gm}
\end{figure*}

To compare structural drift behaviors, consider also the Even Process. Similar
in form to the Golden Mean Process, the Even Process produces populations in
which blocks of consecutive $1$s must be even in length when bounded by $0$s
\cite{Crutchfield2003}. Figure \ref{figure:drift_multiple} compares the drift
of $\Prob[\textsc{Heads}]$ for a single realization of the Biased Coin, Golden
Mean, and Even Processes. One observes that the Even and Biased Coin Processes
reach stasis as the Fixed Coin Process, while the Golden Mean Process reaches
stasis as the Alternating Process. For different realizations, the Even and
Golden Mean Processes might instead reach different stasis points.

It should be noted that the memoryful Golden Mean and Even Processes reach
stasis markedly faster than the memoryless Biased Coin. While Fig.
\ref{figure:drift_multiple} shows only a single realization of each sampling
process type, the top panel of Fig. \ref{figure:tts_subspaces} shows the large
disparity in stasis times holds across all settings of each process's initial
bias. This is one of our first general observations about memoryful processes:
The structure of memoryful processes substantially impacts the average time to
stasis by increasing variance between generations. In the cases shown, time to
stasis is greatly shortened.

\section{Isostructural Subspaces}

\subsection{Subspace Diffusion}

To illustrate the richness of structural drift and to understand how it
affects average time to stasis, we examine the complexity-entropy (CE) diagram
\cite{Feldman2008} of the \eMs\ produced over several realizations of an
arbitrary sampling process. The CE diagram displays how the allelic entropy
$\hmu$ of an \eM\ varies with the allelic complexity $\Cmu$ of its causal
states:
\begin{align}
\Cmu = - \sum_{\causalstate \in \CausalStateSet}
  \Prob(\causalstate) \log_2 \Prob(\causalstate) ~,
\end{align}
where the units are [bits]. The allelic complexity is the Shannon entropy over
an \eM's stationary state distribution $\Prob(\CausalState)$. It measures the
memory needed to maintain the internal state while producing stochastic
outputs. \EM\ minimality guarantees that $\Cmu$ is the smallest amount of
memory required to do so. Since there is a one-to-one correspondence between
processes and their \eMs, a CE diagram is a projection of process space onto
the two coordinates \mbox{$(\hmu,\Cmu)$}. Used in tandem, these two properties
differentiate many types of sampling process, capturing both their intrinsic
memory ($\Cmu$) and the diversity ($\hmu$) of populations they generate.

Two such CE diagrams are shown in Fig. \ref{figure:ce_gm}, illustrating
different subspaces and stasis points reachable by the Golden Mean Process
during structural drift. Consider the left panel first. An \eM\ reaches stasis
by transforming into either the Fixed Coin or the Alternating Process. To reach
the former, the \eM\ begins on the upper curve in the left panel and drifts
until the \mbox{$A \to B$} transition probability nears zero and the inference
algorithm decides to merge states in the next generation. This forces the \eM\
to jump to the Biased Coin subspace on the line \mbox{$\Cmu = 0$} where it will
most likely diffuse until the Fixed Coin stasis point at \mbox{$(\hmu, \Cmu) =
(0, 0)$} is reached. If instead the \mbox{$A \to B$} transition probability
drifts towards zero, the Golden Mean stays on the upper curve until reaching
the Alternating Process stasis point at \mbox{$(\hmu, \Cmu) = (0, 1)$}. Thus,
the two stasis points are differentiated not by $\hmu$ but by $\Cmu$, with the
Alternating Process requiring 1 bit of memory to track its internal state and
the Biased Coin Process requiring none.

What emerges from these diagrams is a broader view of how population structure
drifts in process space. Roughly, the $\eMachine_t$ diffuse locally in the
parameter space specified by the current, fixed architecture of states and
transitions. During this, transition probability estimates vary stochastically
due to sampling variance. Since $\Cmu$ and~$\hmu$ are continuous functions of
the transition probabilities, this variance causes the $\eMachine_t$ to fall
on well defined curves or regions corresponding to a particular process
subspace. (See Figs. 4 and 5 in Ref. \cite{Feldman2008} and the theory for
these curves and regions there.)

We refer to these curves as \emph{isostructural curves} and the associated sets
of \eMs\ as \emph{isostructural subspaces}. They are metastable subspaces of
sampling processes that are quasi-invariant under the structural drift dynamic.
When one or more \eM\ parameters diffuse sufficiently so that inference is
forced to change topology by adding or removing states or transitions to
reflect the statistics of the sample, this quasi-invariance is broken. We call
such topological shifts \emph{subspace jumps} to reflect the new region
occupied by the resulting \eM\ in process space, as visualized by the CE
diagram. Movement between subspaces is often not bidirectional---innovations
from a previous topology may be lost either temporarily (when the innovation
can be restored by returning to the subspace) or permanently. For example, the
Golden Mean subspace commonly jumps to the Biased Coin subspace but the
opposite is highly improbable without mutation. (We consider the latter type of
structured drift elsewhere.)

Before describing the diversity seen in the CE diagram of
Fig. \ref{figure:ce_gm}'s right panel, we first turn to analyze in some
detail the time-to-stasis underlying the behavior illustrated in the left panel.

\subsection{Subspace Decomposition}

\begin{figure}
\includegraphics{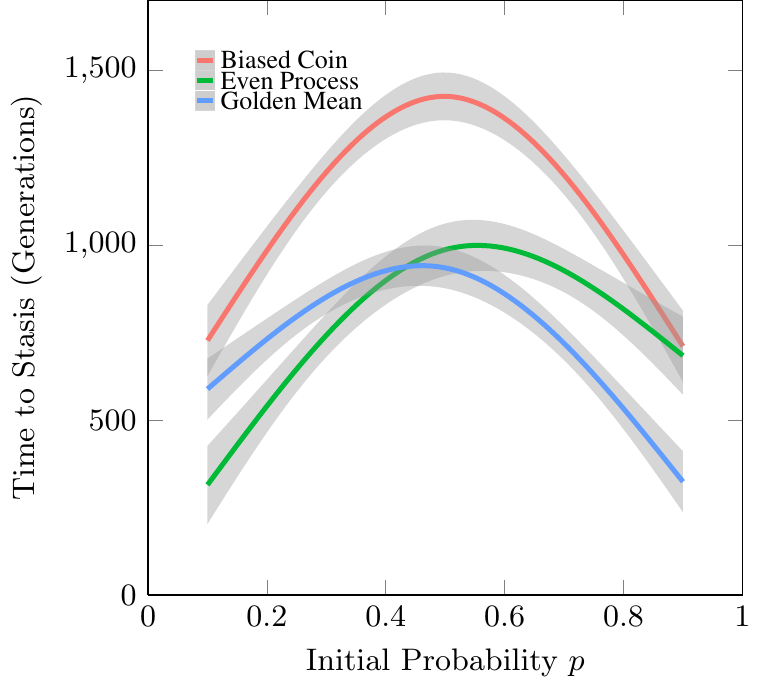}
\vspace{0.5\baselineskip}
\includegraphics{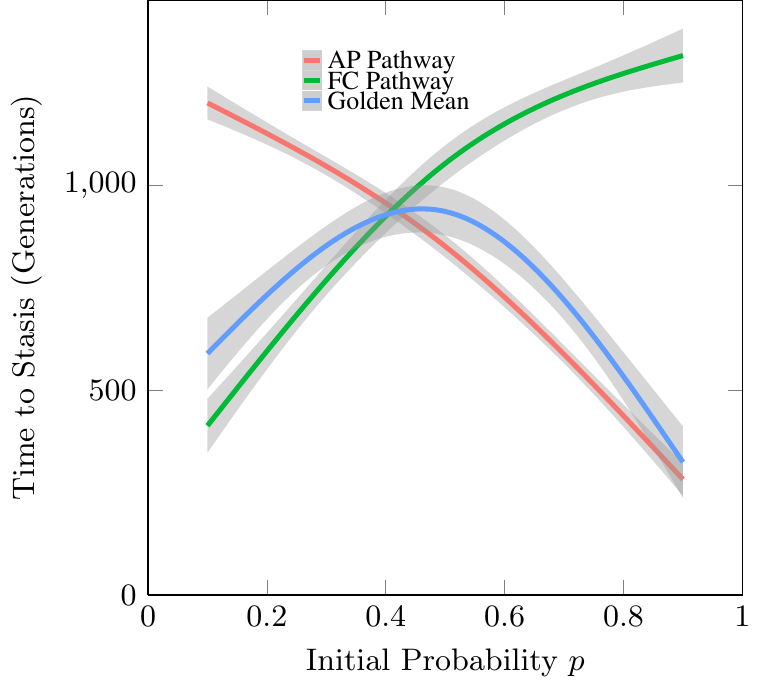}
\vspace{0.5\baselineskip}
\includegraphics{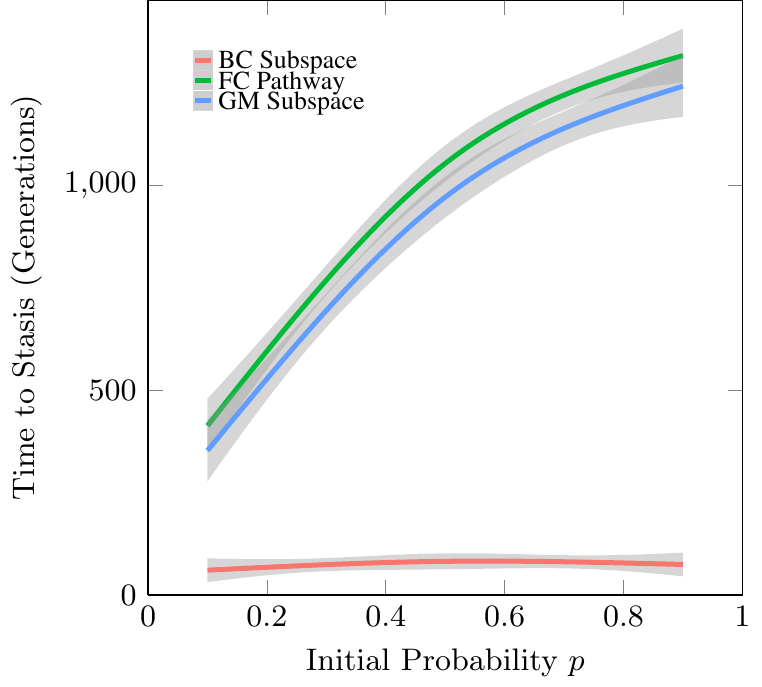}
\caption{
\emph{Top:} Time to stasis of the Golden Mean, Even, and Biased Coin
Processes.
\emph{Middle}: Stasis time of the Golden Mean Process as the weighted sum of
stasis times for the Fixed Coin (FC) and Alternating Process (AP) pathways.
\emph{Bottom}: Stasis time of the FC pathway as the weighted sum of Golden
Mean (GM) and Biased Coin (BC) subspace diffusion times.
}
\label{figure:tts_subspaces}
\end{figure}

A \emph{pathway} is a set of subspaces passed through by any drift realization
starting from some initial process and reaching a specific stasis point. The
time to stasis of a drift process $\mathcal{P}$ is the sum of time spent in
the subspaces~$\gamma$ visited by its pathways to stasis $\rho$, weighted by
the probabilities that these pathways and subspaces will be reached. The time
spent in a subspace $\gamma_{i+1}$ only depends on the transition
parameter(s) of the \eM\ at the time of entry and is otherwise independent of
the prior subspace $\gamma_i$. Thus, calculating the stasis time of a
structured population can be broken down into independent subspace times when
we know the values of the transition parameters at subspace jumps. These
values can be derived both empirically and analytically, and we aim to develop
the latter for general drift processes in future work.

More formally, the time-to-stasis $t_s$ of a drift process~$\mathcal{P}$ is
simply the weighted sum of the stasis times for its connected pathways $\rho$:
\begin{align}
t_s(\mathcal{P}) =
  \sum_{i=1}^{|\rho|} \Prob(\rho_i|\mathcal{P}) t_s(\rho_i|\mathcal{P}) ~,
\end{align}
Similarly, the stasis time of a particular pathway decomposes into the time
spent diffusing in its connected subspaces $\gamma$:
\begin{align}
t_s(\rho_i|\mathcal{P}) = \sum_{i=1}^{|\gamma|}
  \Prob(\gamma_i|\rho_i,\mathcal{P}) t(\gamma_i|\rho_i,\mathcal{P}) ~.
\end{align}
To demonstrate, Fig. \ref{figure:tts_subspaces} shows the stasis time of the
Golden Mean Process (GMP) with initial bias $p_0$ in more detail. Regression
lines along with their 95\% confidence intervals are displayed for simulations
with initial biases $p_0 \in \{0.1, 0.2, \dots, 0.9\}$. The middle panel shows
the total time-to-stasis as the weighted sum of its Fixed Coin (FC) and
Alternating Process (AP) pathways:
\begin{align*}
t_s(\GMP(p_0)) = &\Prob(\FC|\GMP(p_0)) t_s(\FC|\GMP(p_0)) \\
+ &\Prob(\AP|\GMP(p_0)) t_s(\AP|\GMP(p_0)) ~.
\end{align*}
For low $p_0$, the transition from state $A$ to state $B$ is unlikely, so $0$s
are rare and the AP pathway is reached infrequently. Thus, the total stasis
time is initially dominated by the FC pathway ($\Prob(\FC|\GMP(p_0))$ is high).
As \mbox{$p_0 \to 0.3$} and above, the AP pathway is reached more frequently
($\Prob(\AP|\GMP(p_0))$ grows) and its stasis time
begins to influence the total. The FC pathway is less likely as \mbox{$p_0
\to 0.6$} and the total time becomes dominated by the AP pathway
($\Prob(\AP|\GMP(p_0))$ is high).

Since the AP pathway visits only one subspace, the bottom panel shows the
stasis time of the FC pathway as the weighted sum of the Golden Mean (GM) and
Biased Coin (BC) subspace times:
\begin{eqnarray}
&& t_s(\FC|\GMP(p_0)) = \nonumber\\
&& \quad \Prob(\GM|\FC,\GMP(p_0)) t(\GM|\FC,\GMP(p_0)) + \nonumber\\
&& \quad \Prob(\BC|\FC,\GMP(p_0)) t(\BC|\FC,\GMP(p_0)) ~.
\end{eqnarray}
This corresponds to time spent diffusing in the GM subspace \emph{before} the
subspace jump and time spent diffusing in the BC subspace \emph{after} the
subspace jump. Note that the times quoted are simply diffusion times within
a subspace, since not every subspace in a pathway contains a stasis point.

These expressions emphasize the dependence of stasis time on the transition
parameters at jump points as well as on the architecture of isostructural
subspaces in drift process space. For example, if the GM jumps to the BC
subspace at $p = 0.5$, the stasis time will be large since the \eM\ is
maximally far from either stasis point. However, the inference algorithm will
typically jump at very low values of $p$ resulting in a small average stasis
time for the BC subspace in the FC pathway. Due to this, calculating the stasis
time for the GMP requires knowing the AP and FC pathways as well as the value
of $p$ where the GM $\to$ BC jump occurs.

\subsection{Structural Innovation and Loss}

Inference of \eMs\ from finite populations is computationally expensive,
particularly in a sequential setting with many realizations. The topology of
the \eM\ is inferred directly from the statistics of finite samples; both
states and transitions are added and removed over time to capture innovation
and loss of population structure. In the spirit of Kimura's
\emph{pseudo-sampling variable} method \cite{Kimura1980}, we introduce a
pseudo-drift algorithm for efficient drift simulation and increased control of
the trade-off between structural innovation and loss.

Instead of inferring and re-inferring an \eM\ each generation, we explicitly
define the conditions for topological changes to the \eM\ of the previous
generation. To test for \emph{structural innovation}, a random causal state
from the current $\eMachine_t$ is cloned and random incoming transitions are
routed instead to the cloned state. This creates a new model $\eMachine_t'$
that describes the same process. Gaussian noise is then added to the cloned
state's outgoing transitions to represent change in population structure.
The likelihood of the population $\alpha_t^N$ is calculated for both
$\eMachine_t$ and $\eMachine_t'$ and the model with the maximum a posteriori
(MAP) likelihood is retained:
\begin{align}
\eMachine_{MAP} =
  \argmax \{\Prob(\alpha_t^N|\eMachine_t), \Prob(\alpha_t^N|\eMachine_t')\} ~.
\end{align}
If the original $\eMachine_t$ was retained, its transition parameters are
updated by feeding the sample through the model to obtain edge counts which
are normalized to obtain probabilities. This produces a generator for
the next generation's population in a way that allows for innovation. As well,
it side-steps the computational cost of the inference algorithm.

To capture structural loss, we monitor near-zero transition probabilities where
the \eM\ inference algorithm would merge states. When such a transition exists
we test for structural simplification by considering all pairwise mergings of
causal states and select the topology via the MAP likelihood. However, unlike
above, we penalize likelihood using the Akaike Information Criterion
(AIC)~\cite{Akaike1974}:
\begin{align}
\mathrm{AIC} = 2k - 2 \ln(L) ~,
\end{align}
and, in particular, the AIC corrected for finite sample
sizes~\cite{Burnham2002}:
\begin{align}
\mathrm{AICc} = \mathrm{AIC} + \frac{2k(k + 1)}{n - k - 1} ~,
\end{align}
where $k$ is the number of model parameters, $L$ is the sample likelihood, and
$n$ is the sample size. A penalized likelihood is necessary because a smaller
\eM\ is more general and cannot fit the data as well. When penalized by model
size, however, a smaller model with sufficient fit to the data may be selected
over a larger, better fitting model. This method allows loss to occur while
again avoiding the expense of the full \eM\ inference algorithm. Extensive
comparisons with several versions of the latter show that the new pseudo-drift
algorithm produces qualitatively the same behavior.

Having explained how the pseudo-drift algorithm introduces structural
innovation and loss we can now describe the drift runs of Fig.
\ref{figure:ce_gm}'s right panel. In contrast to the left panel, structural
innovation was enabled. The immediate result is that the drift process visits a
much wider diversity of isostructural subspaces---sampling processes that are
markedly more complex. \EMs\ with~$8$ or more states are created, some of which
are quite entropic and so produce high sampling variance. Stasis \eMs\ with
periods $3$, $4$, $5$, and $6$ are seen, while only those with periods $1$ and
$2$ are seen in runs without innovation (left panel).

By way of closing this first discussion of structural drift, it should be
emphasized that none of the preceding phenomena occur in the limit of infinite
populations or infinite sample size. The variance due to finite sampling drives
sequential learning, the diffusion through process space, and the jumps between
isostructural subspaces.

\section{Applications and Extensions}

Much of the previous discussion focused on structural drift as a kind of
stochastic process, with examples and behaviors selected to emphasize the role
of structure. Although there was a certain terminological bias toward neutral
evolution theory since the latter provides an entree to analyzing how
structural drift works, our presentation was intentionally general. Motivated
by a variety of potential applications and extensions, we describe these now
and close with several summary remarks on structural drift itself.

\subsection{Emergent Semantics and Learning in Communication Chains}

Let's return to draw parallels with the opening example of the game of
\emph{Telephone} or, more directly, to the sequential inference of temporal
structure in an utterance passed along a serially coupled communication chain.
There appears to be no shortage of related theories of language evolution.
These range from the population dynamics of Ref. \cite{Komarova2003} and the
ecological dynamics of Ref. \cite{Sole2010} to the cataloging of error sources
in human communication \cite{Campbell1958} and recent efforts to understand
cultural evolution as reflecting learning biases
\cite{Griffiths2008,Chater2009}.

By way of contrast, structural drift captures the language-centric notion of
dynamically changing semantics and demonstrates how behavior is driven by
finite-sample fluctuations within a semantically organized subspace. The
symbols and words in the generated strings have a semantics given by the
structure of a subspace's \eM; see Ref. \cite{Crutchfield1992}. A particularly
simple example was identified quite early in the information-theoretic analysis
of natural language: The Golden Mean \eM\ (Fig. \ref{figure:em_gm}) describes
the role of isolated space symbols in written English \cite[Fig.
1]{Miller1958}. Notably, this structure is responsible for the Mandelbrot-Zipf
power-law scaling of word frequencies \cite{Mandelbrot1953,Zipf1965}. More
generally, though, the semantic theory of \eMs\ shows that causal states
provide dynamic contexts for interpretation as individual symbols and words are
recognized. Quantitatively, the allelic complexity $\Cmu(\eMachine_t)$ is the
total amount of semantic content that can be generated by an $\eMachine_t$
\cite{Crutchfield1992}. In this way, shifts in the architecture of the
$\eMachine_t$ during drift correspond to semantic changes. That is, diffusion
within an isostructural subspace corresponds to constant semantics, while jumps
between isostructural subspaces correspond to semantic innovations (or losses).

In the drift behaviors explored above, the $\eMachine_t$ went to stasis
(\mbox{$\hmu = 0$}) corresponding to periodic formal languages. Clearly, such a
long-term condition falls far short as a model of human communication chains.
The resulting communications, though distant from those at the beginning of the
chain, are not periodic. To more closely capture emergent semantics in the
context of sequential language learning, we have extended structural drift to
include mutation and selection. In future work we will use these extensions to
investigate how the former prevents permanent stasis and the latter enables a
preference for intelligible phrases.

\subsection{Cultural Evolution and Iterated Learning}

Extending these observations, the Iterated Learning Model (ILM) of language
evolution~\cite{Smith2003,Kirby2007} is of particular interest. In this model,
a language evolves by repeated production and acquisition by agents under
cultural pressures and the ``poverty of the stimulus'' \cite{Smith2003}. Via
this process, language is effectively forced through a transmission bottleneck
that requires the learning agent to generalize from finite data. This, in turn,
exerts pressure on the language to adapt to the bias of the learner. Thus, in
contrast to traditional views that the human brain evolved to learn language,
ILM suggests that language also adapts to be learnable by the human brain.

ILM incorporates the sequential learning and propagation of error we discuss
here and provides valuable insight into the effects of error and cultural
mutations on the evolution of language for the ``human niche''. There are
various simulation approaches to ILM with both single and multiple agents based
on, for example, neural networks and Bayesian inference, as well as experiments
with human subjects. We suggest that structural drift could also serve as the
basis for single-agent ILM experiments, as found in Swarup et
al.~\cite{Swarup2009}, where populations of alleles in the former are replaced
by linguistic features of the latter. The benefits are compelling: an
information-theoretic framework for quantifying the trade-off between learner
bias and transmission bottleneck pressures, visualization of cultural evolution
via the CE diagram, and decomposition of the time-to-stasis of linguistic
features in terms of isostructural subspaces as presented above.

\subsection{Epochal Evolution}

Beyond applications to knowledge transmission via serial communication
channels, structural drift gives an alternative view of drift processes in
population genetics. In light of new kinds of evolutionary behavior, it
reframes the original questions about underlying mechanisms and extends their
scope to phenomena that exhibit memory in the sampling process or that derive
from structure in populations. Examples of the latter include niche
construction \cite{Odling-Smee2003}, the effects of environmental toxins
\cite{Medina2007}, changes in predation \cite{Tremblay2008}, and
socio-political factors \cite{Kayser2005} where memory lies in the spatial
distribution of populations. In addition to these, several applications to
areas beyond population genetics proper suggest themselves.

An intriguing parallel exists between structural drift and the longstanding
question about the origins of \emph{punctuated equilibrium} \cite{Gould1977}
when modeled as the dynamics of \emph{epochal evolution}
\cite{Nimwegen1999,Crutchfield2003b}. The possibility of evolution's
intermittent progress---long periods of stasis punctuated by rapid
change---dates back to Fisher's demonstration of metastability in drift
processes with multiple alleles \cite{Fisher1930}.

Epochal evolution, though, presented an alternative to the views of
metastability posed by Fisher's model and Wright's adaptive landscapes
\cite{Wright1932}. Within epochal evolutionary theory, equivalence classes of
genotype fitness, called \emph{subbasins}, are connected by fitness-changing
\emph{portals} to other subbasins. A genotype is free to diffuse within its
subbasin via selectively neutral mutations, until an advantageous mutation
drives genotypes through a portal to a higher-fitness subbasin. An increasing
number of genotypes derive from this founder and diffuse in the new subbasin
until another portal to higher fitness is discovered. Thus, the structure of
the subbasin-portal architecture dictates the punctuated dynamics of
evolution.

Given an adaptive system which learns structure by sampling its past
organization, structural drift theory implies that its evolutionary dynamics
are inevitably described by punctuated equilibria. Diffusion in an
isostructural subspace corresponds to a period of structured equilibrium in a
subbasin and subspace jumps correspond to rapid innovation or loss of
organization during the transit of a portal. In this way, structural drift
establishes a connection between evolutionary innovation and structural change,
identifying the conditions for creation or loss of organization. Extending
structural drift to include mutation and selection will provide a theoretical
framework for epochal evolution using any number of structural constraints in a
population.

\subsection{Evolution of Graph-Structured Populations}

We focused primarily on the drift of sequentially ordered populations in which
the generator (an \eM) captured the structure and randomness in that ordering.
We aimed to show that a population's organization plays a crucial role in its
dynamics. This was, however, only one example of the general class of drift
process we have in mind. For example, computational mechanics also describes
structure in spatially extended systems~\cite{Hanson1997,Varn2004}. Given this,
it is straightforward to build a model of drift in geographically distributed
populations that exhibit spatiotemporal structure.

Though they have not tracked the structural complexity embedded in populations
as we have done here, a number of investigations consider various
classes of structured populations. For example, the evolutionary dynamics of
structured populations have been studied using undirected graphs to represent
correlations between individuals. Edge weights $w_{ij}$ between individuals
$i$ and $j$ give the probability that $i$ will replace $j$ with its offspring
when selected to reproduce.

By studying fixation and selection behavior on different types of graphs,
Lieberman et al found that graph structures can sometimes amplify or suppress
the effects of selection, even guaranteeing the fixation of advantageous
mutations \cite{Lieberman2005}. Jain and Krishna \cite{Jain2002} investigated
the evolution of directed graphs and the emergence of self-reinforcing
autocatalytic networks of interaction. They identified the attractors in these
networks and demonstrated a diverse range of behaviors from the creation of
structural complexity to its collapse and permanent loss.

Graph evolution is a model of population structure complementary to that
presented by structural drift. In the latter, \eM\ structure evolves over time
with nodes representing equivalence classes of the distribution of selectively
neutral alleles. Unlike \eMs, the multinomial sampling of individuals in graph
evolution is a memoryless process. A combined approach will allow one to
examine how amplification and suppression of selection and the emergence of
autocatalysis are affected by external influences on the population structure.
For example, this could include how a population uses temporal memory to
maintain desirable properties in anticipation of structural shifts in the
environment. The result would provide a theory for niche construction in which
a nonlinear dynamics of pattern formation spontaneously changes population
structure.

\section{Final Remarks}

The Fisher-Wright model of genetic drift can be viewed as a random walk of coin
biases, a stochastic process that describes generational change in allele
frequencies based on a strong statistical assumption: The sampling process is
memoryless. Here, we developed a generalized structural drift model that adds
memory to the process and examined the consequences of such population sampling
memory.

Memoryful sampling is a substantial departure from modeling evolutionary
processes with unordered populations. Rather than view structural drift as a
replacement for the well understood theory of genetic drift, and given that the
latter is a special case of structurally drifting populations, we propose that
it be seen as a new avenue for theoretical invention. Given its additional ties
to language and cultural evolution, we believe it will provide a novel
perspective on evolution in nonbiological domains, as well.

The representation selected for the population sampling mechanism was the class
of probabilistic finite-state hidden Markov models called \eMs. We discussed
how a sequential chain of \eMs\ inferred and re-inferred from the finite data
they generate parallels the drift of alleles in a finite population, using
otherwise the same assumptions made by the Fisher-Wright model. The
mathematical foundations developed for the latter and its related models
provide a good deal of quantitative, predictive power. Much of this has yet to be
exploited. In concert with this, \eM\ minimality allowed us to monitor
information processing, information storage, and causal architecture during the
drift process. We introduced the information-theoretic notion of structural
stasis to combine the concepts of deletion, fixation, and periodicity for drift
processes. Generally, structural stasis occurs when the population's allelic
entropy vanishes---a quantity one can calculate in closed form due to the \eM\
representation of the sampling process.

We revisited Kimura and Ohta's early results measuring the time to fixation of
drifting alleles and showed that the generalized structural drift process
reproduces these well known results when staying within the memoryless sampling
process subspace. Starting with structured populations outside of that subspace
led the sampling process to exhibit memory effects including structural
innovation and loss, complex transients, and greatly reduced stasis times.

Simulations demonstrated how an \eM\ diffuses through isostructural process
subspaces during sequential learning. The result was a very complex
time-to-stasis dependence on the initial probability parameter---much more
complicated than Kimura's theory describes. Nonetheless, we showed that a
process' time-to-stasis can be decomposed into sums over these independent
subspaces. Moreover, the time spent in an isostructural subspace depends on the
value of the \eM\ probability parameters at the time of entry. This suggests an
extension to Kimura's theory for predicting the time to stasis for each
isostructural component independently. Much of the phenomenological analysis
was facilitated by the global view of drift process space given by the
complexity-entropy diagram.

Drift processes with memory generally describe the evolution of structured
populations without mutation or selection. Nonetheless, we showed that
structure leads to substantially shorter stasis times. This was seen in drifts
starting with the Biased Coin and Golden Mean Processes, where the Golden Mean
jumps into the Biased Coin subspace close to an absorbing state. This suggests
that even without selection, population structure and sampling memory matter in
evolutionary dynamics. The temporal or spatial memory captured by the \eM\ can
be interpreted as nonrandom mating, reducing the effective population size
$N_e$ and, in doing so, increasing sampling variance. It also suggests that
memoryless models restrict sequential learning and overestimate stasis times
for structured populations.

We demonstrated how structural drift---diffusion, structural innovation and
loss---are controlled by the architecture of connected isostructural subspaces.
Many questions remain about these subspaces. What is the degree of
subspace-jump irreversibility? Can we predict the likelihood of these jumps?
What does the phase portrait of a drift process look like? Thus, to better
understand structural drift, we need to analyze the high-level organization of
generalized drift process space.

Fortunately, \eMs\ are in one-to-one correspondence with structured processes
\cite{Johnson2010}. Thus, the preceding question reduces to understanding the
space of \eMs\ and how they can be connected by diffusion processes. Is the
diffusion within each process subspace predicted by Kimura's theory or some
simple variant? We have given preliminary evidence that it does. And so, there
are reasons to be optimistic that in face of the open-ended complexity of
structural drift, a good deal can be predicted analytically. And this, in turn,
will lead to quantitative applications.

\section*{Acknowledgments}

This work was partially supported by the Defense Advanced Research Projects
Agency (DARPA) Physical Intelligence project via subcontract No. 9060-000709.
The views, opinions, and findings contained here are those of the authors and
should not be interpreted as representing the official views or policies,
either expressed or implied, of the DARPA or the Department of Defense.

\bibliography{library,manual}

\end{document}

%% file: cmechabbrev.tex
\DeclareMathOperator*{\argmax}{argmax}

\newcommand{\eM}     {\mbox{$\epsilon$-machine}}
\newcommand{\eMs}    {\mbox{$\epsilon$-machines}}
\newcommand{\EM}     {\mbox{$\epsilon$-Machine}}
\newcommand{\EMs}    {\mbox{$\epsilon$-Machines}}

\newcommand{\MeasAlphabet}	{\mathcal{A}}

\newcommand{\CausalState}	{ \mathcal{S} }

\newcommand{\causalstate}	{ \sigma }
\newcommand{\CausalStateSet}	{ \boldsymbol{\CausalState} }

\newcommand{\Prob}      {\Pr} 

\newcommand{\Cmu}		{C_\mu}
\newcommand{\hmu}		{h_\mu}

\newcommand{\forward}{+}
\newcommand{\reverse}{-}
\newcommand{\forwardreverse}{\pm} 

\newcommand{\FutureCausalState}	{ {\CausalState}^{\forward} }

\newcommand{\PastCausalState}	{ {\CausalState}^{\reverse} }

\newcommand{\eMachine}	{ M }

\newcommand{\lastindex}[2]{
  \edef\tempa{0}
  \edef\tempb{#2}
  \ifx\tempa\tempb
    \edef\tempc{#1}
  \else
    \edef\tempa{0}
    \edef\tempb{#1}
    \ifx\tempa\tempb
      \edef\tempc{#2}
    \else
      \edef\tempc{#1+#2}
    \fi
  \fi
  \tempc
}

\newcommand{\CSjoint}[1][,]{
   \edef\tempa{:}
   \edef\tempb{#1}
   \ifx\tempa\tempb
      \ensuremath{\FutureCausalState\!#1\PastCausalState}
   \else
      \ensuremath{\FutureCausalState#1\PastCausalState}
   \fi
}
\newcommand{\CSjointKL}[3][,]{
   \edef\tempa{:}
   \edef\tempb{#1}
   \ifx\tempa\tempb
      \ensuremath{\FutureCausalState_{#2}\!#1\PastCausalState_{#3}}
   \else
      \ensuremath{\FutureCausalState_{#2}#1\PastCausalState_{#3}}
   \fi
}

\newif\ifpm
\edef\tempa{\forwardreverse}
\edef\tempb{\pm}
\ifx\tempa\tempb
   \pmfalse
\else
   \pmtrue
\fi


%% file: tikz-prerender.tex
\usetikzlibrary{external}
\tikzsetexternalprefix{figures/}

%% file: theorems.tex
\theoremstyle{plain} 
\theoremstyle{plain} 
\theoremstyle{plain} 
\theoremstyle{plain} 
\theoremstyle{plain} 
\theoremstyle{plain} 
\theoremstyle{plain} \newtheorem{Prop}{Proposition}
\theoremstyle{plain} \newtheorem*{ProProp}{Proof}
\theoremstyle{plain} 
\theoremstyle{plain} 
\theoremstyle{plain} \newtheorem*{Def}{Definition}
\theoremstyle{plain} 

%% file: sdpdsl.bbl
\begin{thebibliography}{10}

\bibitem{Smith1988}
C.~U.~M. Smith, ``{Send reinforcements we're going to advance},'' {\em Biology
  and Philosophy}, vol.~3, no.~2, pp.~214--217, 1988.

\bibitem{Crutchfield1989}
J.~P. Crutchfield and K.~Young, ``{Inferring Statistical Complexity},'' {\em
  Physical Review Letters}, vol.~63, no.~2, pp.~105--108, 1989.

\bibitem{Crutchfield1992}
J.~P. Crutchfield, ``{Semantics and Thermodynamics},'' in {\em Nonlinear
  Modeling and Forecasting} (M.~Casdagli and S.~Eubank, eds.), pp.~317--359,
  Addison-Wesley, 1992.

\bibitem{Shalizi2001a}
C.~R. Shalizi and J.~P. Crutchfield, ``{Computational Mechanics: Pattern and
  Prediction, Structure and Simplicity},'' {\em Journal of Statistical
  Physics}, vol.~104, no.~3, pp.~817--879, 2001.

\bibitem{Kimura1969}
M.~Kimura and T.~Ohta, ``{The Average Number of Generations until Fixation of a
  Mutant Gene in a Finite Population.},'' {\em Genetics}, vol.~61, no.~3,
  pp.~763--771, 1969.

\bibitem{VanNimwegen1999}
E.~van Nimwegen, J.~P. Crutchfield, and M.~Huynen, ``{Neutral evolution of
  mutational robustness},'' {\em Proceedings of the National Academy of
  Sciences of the United States of America}, vol.~96, no.~17, pp.~9716--9720,
  1999.

\bibitem{Bloom2006}
J.~D. Bloom, S.~T. Labthavikul, C.~R. Otey, and F.~H. Arnold, ``{Protein
  stability promotes evolvability},'' {\em Proceedings of the National Academy
  of Sciences of the United States of America}, vol.~103, no.~15,
  pp.~5869--5874, 2006.

\bibitem{Raval2007}
A.~Raval, ``{Molecular Clock on a Neutral Network},'' {\em Physical Review
  Letters}, vol.~99, no.~13, pp.~138104--138108, 2007.

\bibitem{Crutchfield2003a}
J.~P. Crutchfield and P.~K. Schuster, {\em {Evolutionary Dynamics: Exploring
  the Interplay of Selection, Accident, Neutrality, and Function}}.
\newblock Santa Fe Institute Series in the Sciences of Complexity, Oxford
  University Press, 2003.

\bibitem{Koelle2006}
K.~Koelle, S.~Cobey, B.~Grenfell, and M.~Pascual, ``{Epochal evolution shapes
  the phylodynamics of interpandemic influenza A (H3N2) in humans},'' {\em
  Science}, vol.~314, no.~5807, pp.~1898--1903, 2006.

\bibitem{Kimura1983}
M.~Kimura, {\em {The Neutral Theory of Molecular Evolution}}.
\newblock Cambridge University Press, 1983.

\bibitem{Wright1931}
S.~Wright, ``{Evolution in Mendelian Populations},'' {\em Genetics}, vol.~16,
  pp.~97--126, 1931.

\bibitem{Fisher1930}
R.~A. Fisher, {\em {The Genetical Theory of Natural Selection}}.
\newblock Oxford University Press, 1930.

\bibitem{Holsinger2009}
K.~E. Holsinger and B.~S. Weir, ``{Genetics in geographically structured
  populations: Defining, estimating and interpreting $F_{ST}$},'' {\em Nature
  Reviews Genetics}, vol.~10, no.~9, pp.~639--650, 2009.

\bibitem{Gillespie2000}
J.~H. Gillespie, ``{Genetic Drift in an Infinite Population: The
  Pseudohitchhiking Model},'' {\em Genetics}, vol.~155, no.~2, pp.~909--919,
  2000.

\bibitem{Mendel1925}
G.~Mendel, {\em {Experiments in Plant Hybridisation}}.
\newblock Harvard University Press, 1925.

\bibitem{Feller1968}
W.~Feller, {\em {An Introduction to Probability Theory and Its Applications,
  Volume 1}}.
\newblock Wiley, 3rd~ed., 1968.

\bibitem{Gillespie2004}
J.~H. Gillespie, {\em {Population Genetics: A Concise Guide}}.
\newblock Johns Hopkins University Press, 2nd~ed., 2004.

\bibitem{Leibler2010}
S.~Leibler and E.~Kussell, ``{Individual histories and selection in
  heterogeneous populations},'' {\em Proceedings of the National Academy of
  Sciences of the United States of America}, vol.~107, no.~29,
  pp.~13183--13188, 2010.

\bibitem{Pielou1967}
E.~C. Pielou, ``{The use of information theory in the study of the diversity of
  biological populations},'' in {\em Proceedings of the 5th Berkeley Symposium
  on Mathematical Statistics and Probability}, pp.~163--177, 1967.

\bibitem{Crutchfield2003}
J.~P. Crutchfield and D.~P. Feldman, ``{Regularities unseen, randomness
  observed: Levels of entropy convergence},'' {\em CHAOS}, vol.~13, no.~1,
  pp.~25--54, 2003.

\bibitem{Johnson2010}
B.~D. Johnson, J.~P. Crutchfield, C.~J. Ellison, and C.~S. McTague,
  ``Enumerating finitary processes,'' 2010.
\newblock {Santa Fe Institute Working Paper 10-11-027; arXiv:1011.0036v1
  [cs.FL]}.

\bibitem{Feldman2008}
D.~P. Feldman, C.~S. McTague, and J.~P. Crutchfield, ``{The organization of
  intrinsic computation: Complexity-entropy diagrams and the diversity of
  natural information processing},'' {\em CHAOS}, vol.~18, no.~4, pp.~59--73,
  2008.

\bibitem{Kimura1980}
M.~Kimura, ``{Average Time until Fixation of a Mutant Allele in a Finite
  Population under Continued Mutation Pressure: Studies by Analytical,
  Numerical, and Pseudo-Sampling Methods},'' {\em Proceedings of the National
  Academy of Sciences of the United States of America}, vol.~77, no.~1,
  pp.~522--526, 1980.

\bibitem{Akaike1974}
H.~Akaike, ``{A new look at the statistical model identification},'' {\em IEEE
  Transactions on Automatic Control}, vol.~19, no.~6, pp.~716--723, 1974.

\bibitem{Burnham2002}
K.~P. Burnham and D.~Anderson, {\em {Model Selection and Multi-Model
  Inference}}.
\newblock Springer, 2002.

\bibitem{Komarova2003}
N.~Komarova and M.~A. Nowak, ``{Language Dynamics in Finite Populations},''
  {\em Journal of Theoretical Biology}, vol.~221, no.~3, pp.~445--457, 2003.

\bibitem{Sole2010}
R.~V. Sol\'{e}, B.~Corominas-Murtra, and J.~Fortuny, ``{Diversity, competition,
  extinction: The ecophysics of language change},'' {\em Journal of the Royal
  Society Interface}, vol.~7, no.~53, pp.~1647--1664, 2010.

\bibitem{Campbell1958}
D.~T. Campbell, ``{Systematic error on the part of human links in communication
  systems},'' {\em Information and Control}, vol.~1, no.~4, pp.~334--369, 1958.

\bibitem{Griffiths2008}
T.~L. Griffiths, M.~L. Kalish, and S.~Lewandowsky, ``{Theoretical and empirical
  evidence for the impact of inductive biases on cultural evolution},'' {\em
  Philosophical Transactions of the Royal Society B}, vol.~363, no.~1509,
  pp.~3503--14, 2008.

\bibitem{Chater2009}
N.~Chater and M.~H. Christiansen, ``{Language Acquisition Meets Language
  Evolution},'' {\em Cognitive Science}, vol.~34, no.~7, pp.~1131--1157, 2009.

\bibitem{Miller1958}
G.~A. Miller, E.~B. Newman, and E.~A. Friedman, ``{Length-Frequency Statistics
  for Written English},'' {\em Information and Control}, vol.~1, no.~4,
  pp.~370--389, 1958.

\bibitem{Mandelbrot1953}
B.~Mandelbrot, ``{An informational theory of the statistical structure of
  languages},'' in {\em Communication Theory} (W.~Jackson, ed.), pp.~486--502,
  London: Butterworths, 1953.

\bibitem{Zipf1965}
G.~K. Zipf, {\em {The Psycho-Biology of Language: An Introduction to Dynamic
  Philology}}.
\newblock MIT Press, 2nd~ed., 1965.

\bibitem{Smith2003}
K.~Smith, S.~Kirby, and H.~Brighton, ``{Iterated learning: A framework for the
  emergence of language},'' {\em Artificial Life}, vol.~9, no.~4, pp.~371--386,
  2003.

\bibitem{Kirby2007}
S.~Kirby, M.~Dowman, and T.~L. Griffiths, ``{Innateness and culture in the
  evolution of language},'' {\em Proceedings of the National Academy of
  Sciences of the United States of America}, vol.~104, no.~12, pp.~5241--5245,
  2007.

\bibitem{Swarup2009}
S.~Swarup and L.~Gasser, ``{The Iterated Classification Game: A New Model of
  the Cultural Transmission of Language},'' {\em Adaptive Behavior}, vol.~17,
  no.~3, pp.~213--235, 2009.

\bibitem{Odling-Smee2003}
F.~J. Odling-Smee, K.~N. Laland, and M.~W. Feldman, {\em {Niche Construction:
  The Neglected Process in Evolution}}.
\newblock Princeton University Press, 2003.

\bibitem{Medina2007}
M.~H. Medina, J.~A. Correa, and C.~Barata, ``{Micro-evolution due to pollution:
  Possible consequences for ecosystem responses to toxic stress},'' {\em
  Chemosphere}, vol.~67, no.~11, pp.~2105--2114, 2007.

\bibitem{Tremblay2008}
A.~Tremblay, D.~Lesbarreres, T.~Merritt, C.~Wilson, and J.~Gunn, ``{Genetic
  Structure and Phenotypic Plasticity of Yellow Perch (Perca Flavescens)
  Populations Influenced by Habitat, Predation, and Contamination Gradients},''
  {\em Integrated Environmental Assessment and Management}, vol.~4, no.~2,
  pp.~264--266, 2008.

\bibitem{Kayser2005}
M.~Kayser, O.~Lao, K.~Anslinger, C.~Augustin, G.~Bargel, J.~Edelmann, S.~Elias,
  M.~Heinrich, J.~Henke, L.~Henke, C.~Hohoff, A.~Illing, A.~Jonkisz,
  P.~Kuzniar, A.~Lebioda, R.~Lessig, S.~Lewicki, A.~Maciejewska, D.~M. Monies,
  R.~Pawłowski, M.~Poetsch, D.~Schmid, U.~Schmidt, P.~M. Schneider,
  B.~Stradmann-Bellinghausen, R.~Szibor, R.~Wegener, M.~Wozniak,
  M.~Zoledziewska, L.~Roewer, T.~Dobosz, and R.~Ploski, ``{Significant genetic
  differentiation between Poland and Germany follows present-day political
  borders, as revealed by Y-chromosome analysis.},'' {\em Human Genetics},
  vol.~117, no.~5, pp.~428--443, 2005.

\bibitem{Gould1977}
S.~J. Gould and N.~Eldredge, ``{Punctuated equilibria: The tempo and mode of
  evolution reconsidered},'' {\em Paleobiology}, vol.~3, no.~2, pp.~115--151,
  1977.

\bibitem{Nimwegen1999}
E.~van Nimwegen, J.~P. Crutchfield, and M.~Mitchell, ``{Statistical Dynamics of
  the Royal Road Genetic Algorithm},'' {\em Theoretical Computer Science},
  vol.~229, no.~1-2, pp.~41--102, 1999.

\bibitem{Crutchfield2003b}
J.~P. Crutchfield, ``{When Evolution is Revolution---Origins of Innovation},''
  in {\em Evolutionary Dynamics---Exploring the Interplay of Selection,
  Neutrality, Accident, and Function} (J.~P. Crutchfield and P.~K. Schuster,
  eds.), Santa Fe Institute Series in the Sciences of Complexity, pp.~101--133,
  Oxford University Press, 2003.

\bibitem{Wright1932}
S.~Wright, ``{The roles of mutation, inbreeding, crossbreeding, and selection
  in evolution},'' in {\em Proceedings of the 6th International Congress on
  Genetics}, pp.~355--366, 1932.

\bibitem{Hanson1997}
J.~E. Hanson and J.~P. Crutchfield, ``{Computational Mechanics of Cellular
  Automata: An Example},'' {\em Physica D}, vol.~103, no.~1-4, pp.~169--189,
  1997.

\bibitem{Varn2004}
D.~P. Varn and J.~P. Crutchfield, ``{From Finite to Infinite Range Order via
  Annealing: The Causal Architecture of Deformation Faulting in Annealed
  Close-Packed Crystals},'' {\em Physics Letters A}, vol.~324, no.~4,
  pp.~299--307, 2004.

\bibitem{Lieberman2005}
E.~Lieberman, C.~Hauert, and M.~A. Nowak, ``{Evolutionary dynamics on
  graphs},'' {\em Nature}, vol.~433, no.~7023, pp.~312--316, 2005.

\bibitem{Jain2002}
S.~Jain and S.~Krishna, ``{Graph theory and the evolution of autocatalytic
  networks},'' in {\em Handbook of Graphs and Networks} (S.~Bornholdt and H.~G.
  Schuster, eds.), pp.~355--395, Wiley-VCH Verlag GmbH \& Co. KGaA, 2002.

\end{thebibliography}
